\documentclass{article}

\usepackage{PRIMEarxiv}

\usepackage[utf8]{inputenc} % allow utf-8 input
\usepackage[T1]{fontenc}    % use 8-bit T1 fonts
\usepackage{hyperref}       % hyperlinks
\usepackage{url}            % simple URL typesetting
\usepackage{booktabs}       % professional-quality tables
\usepackage{amsfonts}       % blackboard math symbols
\usepackage{nicefrac}       % compact symbols for 1/2, etc.
\usepackage{microtype}      % microtypography
\usepackage{lipsum}
\usepackage{fancyhdr}       % header
\usepackage{graphicx}       % graphics
\graphicspath{{media/}}     % organize your images and other figures under media/ folder

 \usepackage{array,multirow}
\usepackage{amsmath,amsfonts}
\usepackage{algorithmic}
\usepackage{array}
\usepackage[caption=false,font=normalsize,labelfont=sf,textfont=sf]{subfig}
\usepackage{textcomp}
\usepackage{stfloats}
\usepackage[ruled,vlined]{algorithm2e} 
\usepackage{verbatim}

\title{AI-Enabled Lung Cancer Prognosis
\thanks{
This is the author's version of a book chapter entitled: "Cancer Research: An Interdisciplinary Approach", Springer.\\

M. Darvish and L. Ren are with the University of Maryland School of Medicine, Baltimore, MD 21201. \\
R. Trask, P. Tallon, M. Khansari, and B. Yousefi are with the University of Maryland, College Park, MD 20742. \\
M. Hershman is with the University of Pennsylvania and Boise Radiology Group, Philadelphia, PA 19104. \\

**Corresponding author:\\
B. Yousefi, BSE-4117, 9636 Gudelsky Drive, Rockville, MD 20850, Tel: 240-665-6529, Email: byousefi@umd.edu\\
}\\
}
\author{
  Mahtab Darvish, Ryan Trask, Patrick Tallon, Mélina Khansari, \\
\textbf{Lei Ren, Michelle Hershman, and Bardia Yousefi}
}

\begin{document}
\maketitle

\begin{abstract}
Lung cancer is the primary cause of cancer-related mortality, claiming approximately 1.79 million lives globally in 2020, with an estimated 2.21 million new cases diagnosed within the same period. Among these, Non-Small Cell Lung Cancer (NSCLC) is the predominant subtype, characterized by a notably bleak prognosis and low overall survival rate of approximately 25$\%$ over five years across all disease stages. However, survival outcomes vary considerably based on the stage at diagnosis and the therapeutic interventions administered. Recent advancements in artificial intelligence (AI) have revolutionized the landscape of lung cancer prognosis. AI-driven methodologies, including machine learning and deep learning algorithms, have shown promise in enhancing survival prediction accuracy by efficiently analyzing complex multi-omics data and integrating diverse clinical variables. By leveraging AI techniques, clinicians can harness comprehensive prognostic insights to tailor personalized treatment strategies, ultimately improving patient outcomes in NSCLC. Overviewing AI-driven data processing can significantly help bolster the understanding and provide better directions for using such systems.

\end{abstract}

\keywords{Lung cancer prognosis \and Artificial intelligence in healthcare \and High dimensional data analysis \and Deep learning for cancer prognosis \and Radiomics \and Multiomics.}

\section{Introduction}
NSCLC (non-small cell lung cancer) is the most common type of lung cancer, accounting for 84$\%$ of lung cancer diagnoses ~\cite{rr1}, and number of cases are trended to increase globally ~\cite{rr2}. Radiomics extracted from various imaging modalities, omics, and pathological data have been found to possess significant predictive power for identifying survival responses in lung cancer patients when analyzed using various machine-learning models. CT (computed tomography) and Positron emission tomography (PET) imaging are the most common imaging modalities used for noninvasive treatment planning. Radiomics extracted from these imaging systems significantly aid AI (artificial intelligence) methods to excel in the computer-aided prognosis of cancer and help treatment planning. 

\begin{figure}[t]
\begin{center}
   \includegraphics[width=0.99\linewidth]{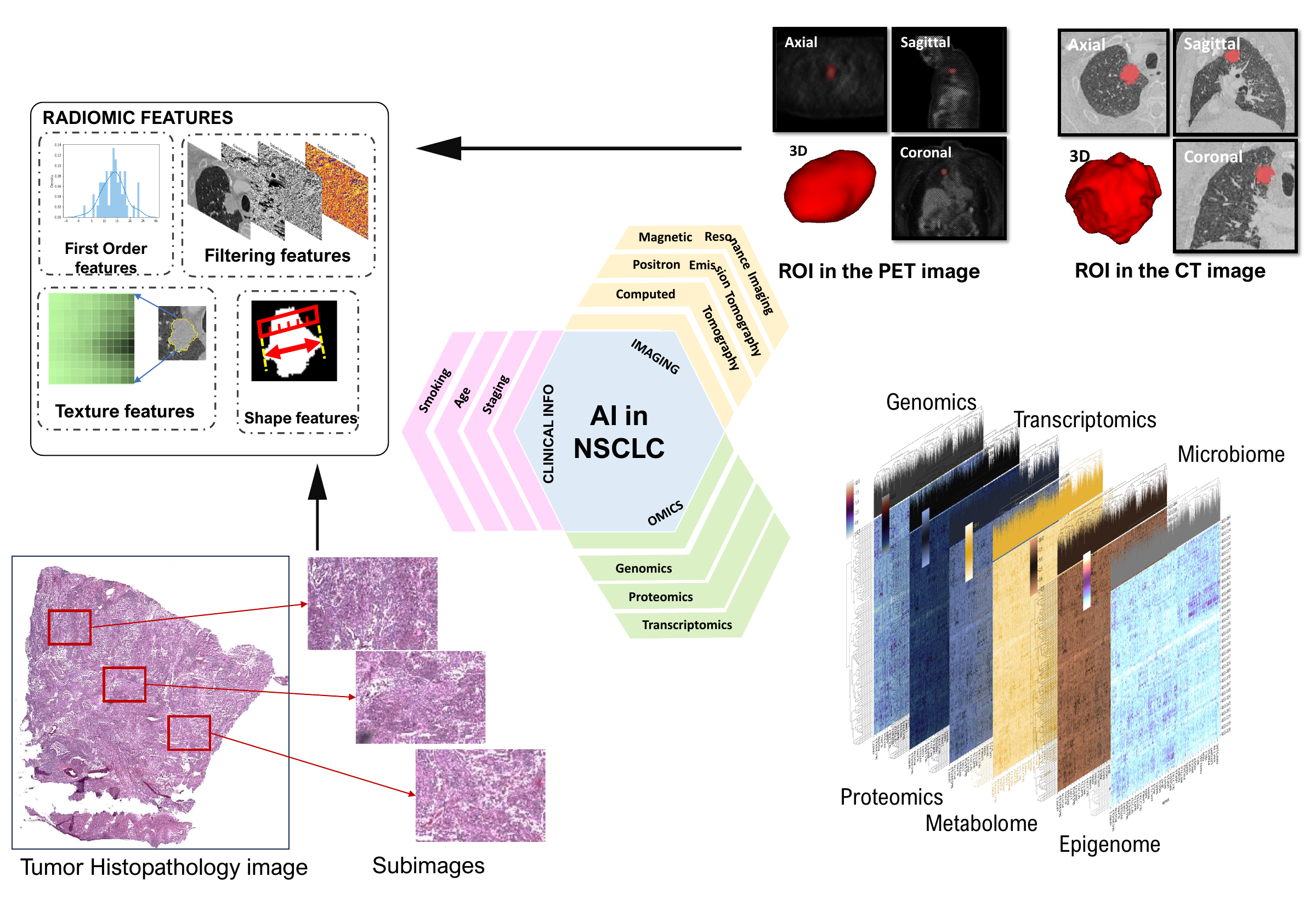}
\end{center}
   \caption{The AI-based NSCLC therapy pipeline shows the major HD data integration involving imaging biomarkers, omics, and clinical information. Imaging biomarkers can be integrated with other data, omics, and clinicals, to improve precision medicine.}
\label{fig1}
\end{figure}

Lung cancer manifests mesoscopic-scale phenotypic traits often imperceptible to the human eye. However, these traits can be non-invasively captured through radiomic features in medical imaging, forming a high-dimensional (HD) data space conducive to machine learning. Leveraging AI, radiomic features enable effective patient risk stratification, prediction of histological and molecular outcomes, and assessment of clinical measures, thus advancing precision medicine in lung cancer care. In contrast to tissue sampling-centric methods, radiomics-based approaches offer advantages in terms of non-invasiveness, reproducibility, cost-effectiveness, and reduced vulnerability to intra-tumoral heterogeneity. Despite the grand ability of such data to encode NSCLC’s characteristics, the large number of covariates can hinder modeling due to overfitting affecting the generalizability of models ~\cite{rr3}. To design such models, there should be great investigation toward all the components involved in data collection, such as biological factors, imaging, pathology, and the AI methodology to tackle HD data processing (Figure \ref{fig1}). This chapter overviews the integration of AI deciphering HD data and their effects on the prognosis of NSCLC and the AI methodologies used to tackle such a problem.

\section{HD gene expression for prognosis of NSCLC}
While traditional staging systems provide valuable clinical information, they often fall short of capturing the unique molecular portrait of each tumor. Harnessing the HD gene expression has emerged as a crucial avenue for unraveling prognostic insights. Exploring the profound implications of genetic information in shaping the prognosis of NSCLC patients through the patterns encoded in the genetic data, researchers strive to unveil novel markers and comprehensive signatures that hold the promise of refining prognostic predictions and advancing personalized therapeutic strategies for such formidable diseases. Enter gene expression profiling (GEP), a ground-breaking technology empowering personalized care by decoding the whispers within cancer cells themselves ~\cite{r1}.

Traditional NSCLC staging, relying on tumor size, lymph node involvement, and metastasis, paints a limited picture. It neglects the intricate molecular choreography within tumor cells that dictates individual patient trajectories ~\cite{r2}. GEP steps in, analyzing the activity of thousands of genes within a tumor sample. By illuminating distinct patterns associated with aggressiveness, response to therapy, and ultimately, survival, it offers a deeper understanding of individual tumor behavior ~\cite{r3}.

Landmark studies have demonstrated the superiority of GEP signatures compared to traditional staging. In a groundbreaking publication, Beer et al. identified a 50-gene signature that independently predicted survival in early-stage NSCLC patients, even after accounting for the stage. This work showcased the ability of GEP to uncover hidden prognostic information missed by conventional means, offering hope for more personalized predictions ~\cite{r4}. Similarly, Kratz et al. demonstrated that the incorporation of the molecular prognostic classifier, which integrates the prognostic expression levels of 11 cancer-related target genes, significantly improves risk stratification and survival predictions of NSCLC compared to the conventional staging systems. This study emphasizes the importance of incorporating molecular descriptors of tumor biology into staging systems to achieve more accurate risk stratification and prognosis ~\cite{r5} Välk et al. conducted a screening of 81 NSCLC samples utilizing whole-genome gene expression microarrays to identify differentially expressed genes and novel NSCLC biomarkers. Their analysis revealed upregulated expression of novel genes in NSCLC, including SPAG5, POLH, KIF23, and RAD54L, associated with mitotic spindle formation, DNA repair, chromosome segregation, and dsDNA break repair, respectively.

  % ------------------------------------ Figure 1 --------------------
  % ------------------------------------ Figure 1 --------------------

In addition, they identified several downregulated genes in NSCLC, such as SGCG, NLRC4, MMRN1, and SFTPD, involved in extracellular matrix formation, apoptosis, blood vessel leakage, and inflammation, respectively. A significant correlation was observed between RNA degradation and survival in adenocarcinoma cases. The study demonstrated improved prediction values in a group selection based on molecular profiles compared to histology ~\cite{r6}.  In another study, Bhattacharjee et al. developed a molecular classification for lung carcinoma and analyzed mRNA expression in 186 lung tumor samples, including 139 adenocarcinomas. The analysis revealed distinct subclasses of lung adenocarcinoma, including those with high expression of neuroendocrine genes and type II pneumocyte genes. Adenocarcinomas with neuroendocrine gene expression had poorer outcomes. This classification method could differentiate primary lung adenocarcinomas from metastases of extra-pulmonary origin, suggesting its potential for aiding in lung cancer diagnosis ~\cite{r7}. Additionally, Guo and colleagues devised a computational model to predict individual patient outcomes based on gene expression profiles in lung adenocarcinoma. They identified a 37-gene signature through advanced bioinformatics, evaluating its prognostic power using hierarchical clustering and Kaplan-Meier analysis. Applying the signature to 84 patients yielded a 96$\%$ predictive accuracy, categorizing patients into groups with varying prognoses. Notably, patients with stage I disease predominantly fell into the group with a good prognosis. These findings suggest the potential for personalized therapy guided by prediction models based on a small set of marker genes ~\cite{r8} These examples highlight the potential of GEP to move beyond traditional limitations and guide better clinical decision-making.

The power of GEP extends beyond mere survival prediction. Identifying specific gene expression patterns associated with aggressive behavior or response to different treatment options paves the way for personalized prognostication and treatment decisions. For instance, Kikuchi and colleagues conducted innovative studies aimed at correlating a patient's gene expression profile with the most effective chemotherapy regimen. They analyzed 37 NSCLCs using cDNA microarray analysis and compared gene expression data with NSCLC sensitivity to six anti-cancer drugs. Their findings revealed significant associations between gene expression levels and chemosensitivity. This suggests the potential for identifying predictive markers for chemotherapy agents ~\cite{r9} Additionally, Petty and colleagues examined the genetic profile of NSCLC patients to predict their response to platinum-based chemotherapy. They identified 17 genes associated with treatment responsiveness, particularly highlighting lysosomal protease inhibitors like serpinB3 and cystatin C. Applying their findings to predict the response of additional patients, they achieved 72% accuracy in distinguishing responders from non-responders ~\cite{r10}.
The field of GEP-based prognosis is constantly evolving. Liquid biopsies, analyzing tumor DNA shed into blood, offer a minimally invasive approach to capturing gene expression information, overcoming the limitations of tissue biopsies ~\cite{r11}. Imamura et al. (2020) evaluate the efficacy of monitoring circulating tumor DNA (ctDNA) in EGFR-mutant non-small-cell lung cancer patients treated with epidermal growth factor receptor tyrosine kinase inhibitors. Researchers analyzed serial plasma samples from 52 patients, observing dynamic changes in plasma mutation scores during treatment. The study highlights the usefulness of ctDNA monitoring in assessing treatment responses and driver oncogene status, revealing clonal heterogeneity and the genetic evolution of cancer at an individual level ~\cite{r12}. In a separate study, Pender et al. illustrated the application of droplet digital PCR (ddPCR) in identifying low-frequency mutations in lung cancer, with a specific focus on KRAS mutations commonly found in NSCLC. Their research involved testing on KRAS mutant cell lines as well as DNA obtained from tumor tissue of lung cancer patients, revealing a notable level of sensitivity ~\cite{r13}. These studies show promising results using liquid biopsies to monitor disease progression and response to therapy in NSCLC patients, potentially paving the way for more personalized treatment monitoring and adjustments.

Furthermore, exploring the interplay between GEP and other factors like the immune system or the tumor microenvironment could unlock new avenues for prognostication and therapeutic targeting. Studies like the one conducted by Ye et al. suggest that understanding the interaction between GEP and the immune system could aid in predicting response to newer immunotherapy treatments, offering a glimpse into the future of personalized cancer care ~\cite{r14}. 

Integrating GEP with other prognostic tools like imaging techniques could further refine risk stratification and treatment decision-making. For example, Aerts et al showed that radiomics-based phenotyping could improve the stratification and assessment of response to tyrosine kinase inhibitors (TKIs) in NSCLC patients ~\cite{r15}. In a more recent study by Chen et al., researchers investigated the correlation between radiological imaging, gene expression patterns, and patient outcomes in NSCLC. They conducted an extensive analysis of 116 NSCLC cases, utilizing CT images, gene expression data, and clinical factors. Through meticulous analysis, radiomic and genomic features were extracted and scrutinized, resulting in the creation of risk scores for each patient pertaining to overall survival (OS). Notably, a fusion survival model was developed, integrating CT images, gene expression data, and clinical factors. This model effectively stratified patients into low- and high-risk groups with notable accuracy, surpassing traditional unimodal data-based approaches. The success of this fusion model underscores the potential of integrating diverse datasets for more precise risk assessment and enhanced patient care strategies in managing NSCLC ~\cite{r16}.

\section{Pathological Imaging}
Pathological data serves as a critical cornerstone in understanding the intricacies of NSCLC prognosis. By scrutinizing histopathological features and molecular signatures, researchers aim to crack invaluable information into disease progression, treatment response, and patient outcomes. This section delves into the crucial role of pathological data in shaping prognostic models for NSCLC, highlighting its potential to refine risk stratification and guide tailored therapeutic interventions in clinical practice. Employing machine learning on the multiple tissue microarray data for distinguishing short- and long- term survival indicate the potential of automated image features in prognostic predictions for lung cancer ~\cite{r300}, which also suggests that quantitative features derived from histopathology images can enhance survival predictions and improving current practices. Such approaches hold promise for precision oncology, offering insights applicable to organ pathologies.

The development and validation of a pathology image-based predictive model for the prognosis of lung adenocarcinoma (ADC) patients across multiple independent cohorts through the extraction of morphological features from formalin-fixed paraffin-embedded tumor tissues indicated some potential for widespread applicability in ADC prognosis prediction ~\cite{r302}. Imaging features were also used for decoding pathological imaging features and their interactions with the prediction response of NSCLC. For example, three radiomic features related to primary tumor sphericity and lymph node homogeneity were significantly predictive of pathological complete response (pCR), while two features quantifying lymph node homogeneity were predictive of gross residual disease (GRD). These features are used for assessing pathological response after neoadjuvant chemoradiation in locally advanced NSCLC ~\cite{r301}. The combination of radiomic and clinical data outperformed other feature sets, suggesting the importance of lymph node phenotypic information in predicting pathological response. 

A CAD-based HD data processing for assessing PD-L1 expression, crucial for anti-PD-1/PD-L1 treatment in NSCLC, showed a strong correlation with manual pathologist scores, demonstrating high concordance across varying PD-L1 cutoffs and provided efficient and comparable PD-L1 assessment, supporting treatment decisions and assay evaluations. ~\cite{r303}. A combination of data from gene expressions and histopathological imaging features can provide insight into NSCLC prognosis, which suggests that gene expressions exhibit slightly better prognostic performance, and there's a weak correlation between most gene expressions and imaging features ~\cite{r304}. The study also emphasized the importance of understanding the roles of molecular and imaging data in cancer prognosis modeling, highlighting the need for further methodological development and investigation into the interconnections between these data types.

Models trained on H\&E-stained Whole Slide Images (WSIs) of transbronchial lung biopsy (TBLB) specimens may lead to high accuracy in classifying ADC, squamous cell carcinoma (SCC), small-cell lung cancer (SCLC), and non-neoplastic cases ~\cite{r305}. Studies of clinically relevant molecular phenotypes from whole-slide histopathology images using human-interpretable image features (HIFs) suggested better addressing the interpretability challenge in pathology workflow ~\cite{r306}. These HIFs, correlating with known tumor microenvironment markers, exhibit predictive capability for diverse molecular signatures, demonstrating a comprehensive and interpretable insight into the spatial architecture of the tumor microenvironment. Aiming to identify crucial features for classifying subtypes, radiomics may lead to optimized classifier and achieving reliable precision, which demonstrates the impressive potential of radiomics in accurately classifying pathologically confirmed NSCLC subtypes, with implications for treatment planning and precision medicine ~\cite{r307}.

\section{HD Imaging Biomarkers}
\begin{figure}[t]
\begin{center}
   \includegraphics[width=0.99\linewidth]{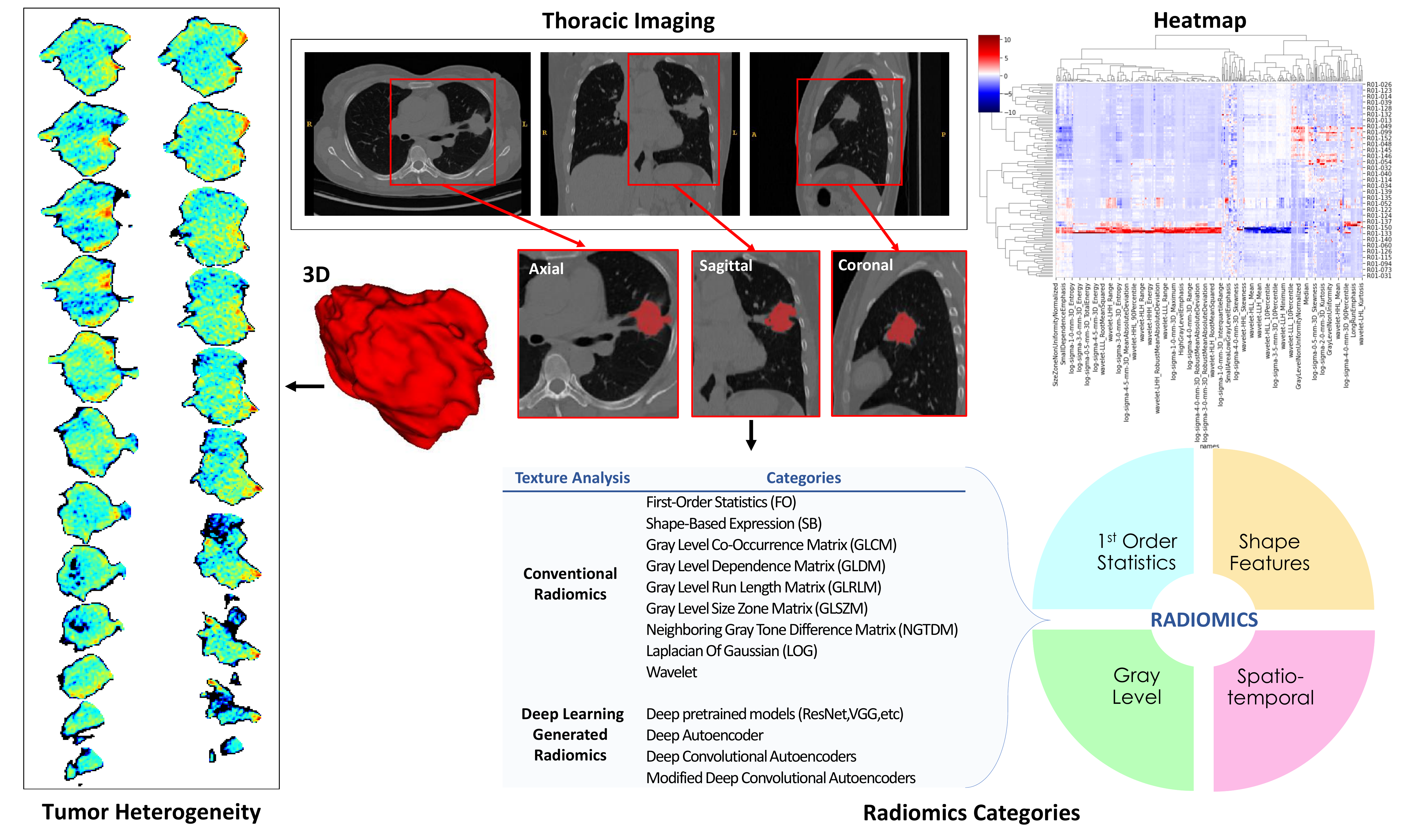}
\end{center}
   \caption{Radiomics involves the extraction and analysis of quantitative features from medical images, enabling the characterization of tumor phenotype and behavior. These features are categorized into four main groups: first-order statistics, which describe voxel intensity distributions and capture spatial relationships between voxel intensities; shape features, which decode the information related to the shape of tumor; and gray level and spatiotemporal features, which encounter texture, pixel intensity, and different features such as Laplacian of Gaussian (LoG), and Wavelet descriptors, respectively, providing insights into tumor heterogeneity and complexity.}
\label{fig2}
\end{figure}

Radiomics involves translating digital medical images into quantitative data, aiming to generate imaging biomarkers for clinical decision support in cancer care. By utilizing imaging data from routine clinical work-ups, radiomics enhances the understanding of tumor biology and supports precision medicine (Figure \ref{fig2}). The noninvasive nature of radiomics enables comprehensive assessments of tumors and their microenvironments, including temporal and spatial heterogeneity. While there's a surge in computational medical imaging publications emphasizing the utility of imaging biomarkers in oncology, this highlights the need for improvement and standardization to enable routine clinical adoption of radiomics as a clinical biomarker ~\cite{r98}. Quantitative features, categorized into intensity, structure, texture/gradient, and wavelet, reveal subvisual attributes correlated with disease pathogenesis. Numerous studies have explored the correlation between these features and the malignant potential of nodules on chest CTs. In cancer patients, these nodules' features also correlate with prognosis and mutation status. However, radiomics faces challenges such as non-standardized acquisition parameters, inconsistent methods, and limited reproducibility. Ongoing research aims to address these limitations, enhancing the acceptance of radiomics within the medical community ~\cite{r99}. Dealing with high dimensional radiomics is also considered to be a challenge in AI in NSCLC prognosis using imaging data similar to genomic applications. While a combination of both, Radiogenomics, is commonly used to ensure an optimized strategy to discover prognostic imaging biomarkers. By correlating image features with coexpressed gene clusters, predictive models were built, achieving significant correlations and accurate predictions ~\cite{r51}. Notably, tumor size, edge shape, and sharpness emerged as top prognostic indicators, suggesting the potential of this radiogenomics approach to expedite the assessment of new imaging modalities in personalized medicine for NSCLC.

Relationship between CT-based radiomic features and epidermal growth factor receptor (EGFR) mutation status was another innovative approach that has been used in many studies such as; exploring 11 radiomics from different categories associated with EGFR mutation in 298 patients with surgically resected peripheral lung adenocarcinomas ~\cite{r102}; 485 radiomics from pre-therapy CT scans associated with EGFR mutation suggesting the potential of radiomic features in guiding treatment decisions, particularly for EGFR tyrosine kinase inhibitor (TKI) therapy ~\cite{r113}; EGFR and KRAS mutations status connected to radiomics also revealed a potential correlation between EGFR mutation status and CT scan imaging phenotypes ~\cite{r117}; radiomics associated with EGFR-targeted therapy in advanced NSCLC patients with EGFR mutations, circulating-tumor DNA (ctDNA), and clinical factors which led to predicting progression-free survival through dealing with HD data ~\cite{r118,r119}. These methods are involved with HD data processing and finding the imaging phenotypes and then their association with genomics and clinical factors. This signifies the importance of dealing with HD imaging biomarkers and their value within the application of AI for NSCLC prognosis. Besides radiogenomics, radiopathomic data also has significant popularity, where radiomic and pathomic features are fused to provide performance status ~\cite{r99+1}. 

The correlation of clinicopathological (CP) and imaging data with 1-, 3-, and 5-year overall survival (OS) of surgically treated patients, both alone and combined, revealed that combining CP and imaging parameters yielded the best predictive values. Normalization influenced the performance of certain imaging features in predicting OS, indicating the potential of density correction for optimizing predictive models ~\cite{r116, r307}. Normalization, or known as harmonization, of radiomics showed its strength in helping to predict patients’ OS. The Nested ComBat ~\cite{r121} and OPNested ComBat ~\cite{r122} methods allow for harmonization by multiple imaging parameters, while the GMM ComBat method uses a Gaussian Mixture Model for grouping scans based on distribution shape before harmonization. The findings suggest promise for improved standardization and generalizability in datasets with multiple or unknown imaging parameters.

  % ------------------------------------ Figure 2 --------------------
  % ------------------------------------ Figure 2 --------------------

Radiomics-based features showing potential for a computational model aiding histopathological diagnosis and metastatic prediction in lung cancer. Such studies offer a "virtual biopsy" without requiring whole-body imaging scanning, providing valuable support for therapy decision-making through machine learning to predict lung cancer histopathology and metastases ~\cite{r114}. Notably, quantitative CT analysis from surgically resected lung adenocarcinoma (ADC), integrating various features, proved effective in noninvasively identifying the micropapillary component. The findings emphasize the diagnostic value of combining radiomic features with clinical parameters for improved treatment planning in lung ADC ~\cite{r105}. Computational methods integrating computerized subtyping and prognosis for NSCLC suggest that quantitative analysis can effectively represent tissue characteristics, offering a viable alternative to invasive pathological methods ~\cite{r108}. 
The computer-aided detection (CAD) system for an integrated framework involving cell detection, segmentation, classification, image marker discovery, and survival analysis is also an interesting process that utilizes a robust cell segmentation algorithm and various classification techniques, such as random forest and AdaBoost ~\cite{r100}. Such models acquiring correlations between discovered image markers and NSCLC subtypes, prognostic image markers related to staining characteristics, and nuclear inhomogeneity significantly affect the prediction of NSCLC patients' survival. 

Through integrating immunohistochemical staining, GLASS-AI, a machine learning-based histological image analysis tool, identified dysregulation of Mapk/Erk signaling in high-grade lung adenocarcinomas and locally advanced tumor regions ~\cite{r123}. This underscores the utility of GLASS-AI in preclinical models and the synergy of machine learning with molecular biology techniques for unraveling the molecular pathways in cancer progression. Such studies are used for HD radiomics-driven addressing interobserver variability in manual segmentation of NSCLC ~\cite{r120}.
Interaction of radiomics and demographics is also commonly used due to the importance of these attributes ~\cite{r99+2}, ~\cite{r99+3}, ~\cite{r110,r111,r112}. Some findings suggest that baseline radiomics of lung cancer screening CT scans can effectively assess the risk of cancer development ~\cite{r101}, radiomics outperforms other approaches, such as the Lung Imaging Reporting and Data System, volume-only methods, and the McWilliams risk assessment model. In addition, the effectiveness of CT radiomic features, in conjunction with clinical data, for diagnosing distant metastasis in lung cancer was shown by analyzing the predictive potential of CT radiomic features for distant metastasis in lung cancer ~\cite{r109}.
Some results showed that both 2D and 3D CT radiomics features demonstrated prognostic ability through comparison of the prognostic performance of 2D and 3D radiomics features in CT images of NSCLC, but 2D features exhibited better performance in the study, suggesting their more favorable use considering the cost of feature calculation ~\cite{r103}. However, combining optimized radiomics signatures from 2D and/or 3D CT images with clinical predictors showed superior prognostic performance ~\cite{r115}.

The radiomic features’ dynamics in NSCLC during therapy and their impact on prognostic models were found to significantly change during radiation therapy ~\cite{r104}. This demonstrates that pretreatment compactness enhances overall survival and distant metastase prediction. Notably, texture strength at the end of treatment effectively stratifies patients for local recurrence risk. The findings indicate that radiomics features, altered by radiation therapy, can serve as indicators of tumor response, and contribute to prognostic models. Moreover, texture features, specifically the standard deviation and mean value of positive pixels, were found to be associated with tumor hypoxia (measured by Glut1/pimonidazole) and angiogenesis (evaluated using CD34) ~\cite{r106}.

Multimodal radiomics, extracting diverse features from medical images through bioinformatic methods, holds promise in predicting tumor biology and behavior. Particularly in 18F-fluorodeoxyglucose positron emission tomography (PET) plus CT radiomics, early data suggest improved tumor volume definition and prediction of radiation toxicity and treatment response in radiation therapy ~\cite{r199}. The impact of respiratory motion on image features investigation involving conventional and respiratory-gated PET/CT images of lung cancer patients suggested that certain features, such as sphericity and entropy for PET, and minimum intensity and RMS for CT, show minimal variability ~\cite{r200}. The association of multimodal imaging features with gene expression pathways, such as hypoxia and the KRAS pathway, was shown through exploring PET/CT imaging for NSCLC prognosis ~\cite{r201}.
AI methods to process imaging biomarkers are often involved with machine learning techniques. Different strategies to handle HD data can provide different outcomes which entails projecting HD imaging biomarkers manifold to low-dimensional (LD) space to preserve characteristics of datasets. To tackle challenges like feature redundancy and unbalanced data, the study employed techniques such as Principal Component Analysis ~\cite{r107,r118,r119,r120, r121,r122}, SMOTE ~\cite{r107}, Clustering ~\cite{r118,r119}, and LASSO ~\cite{r115,r201}.

\section{Deep Learning Integrated Prognosis}
\begin{figure}[t]
\begin{center}
   \includegraphics[width=0.99\linewidth]{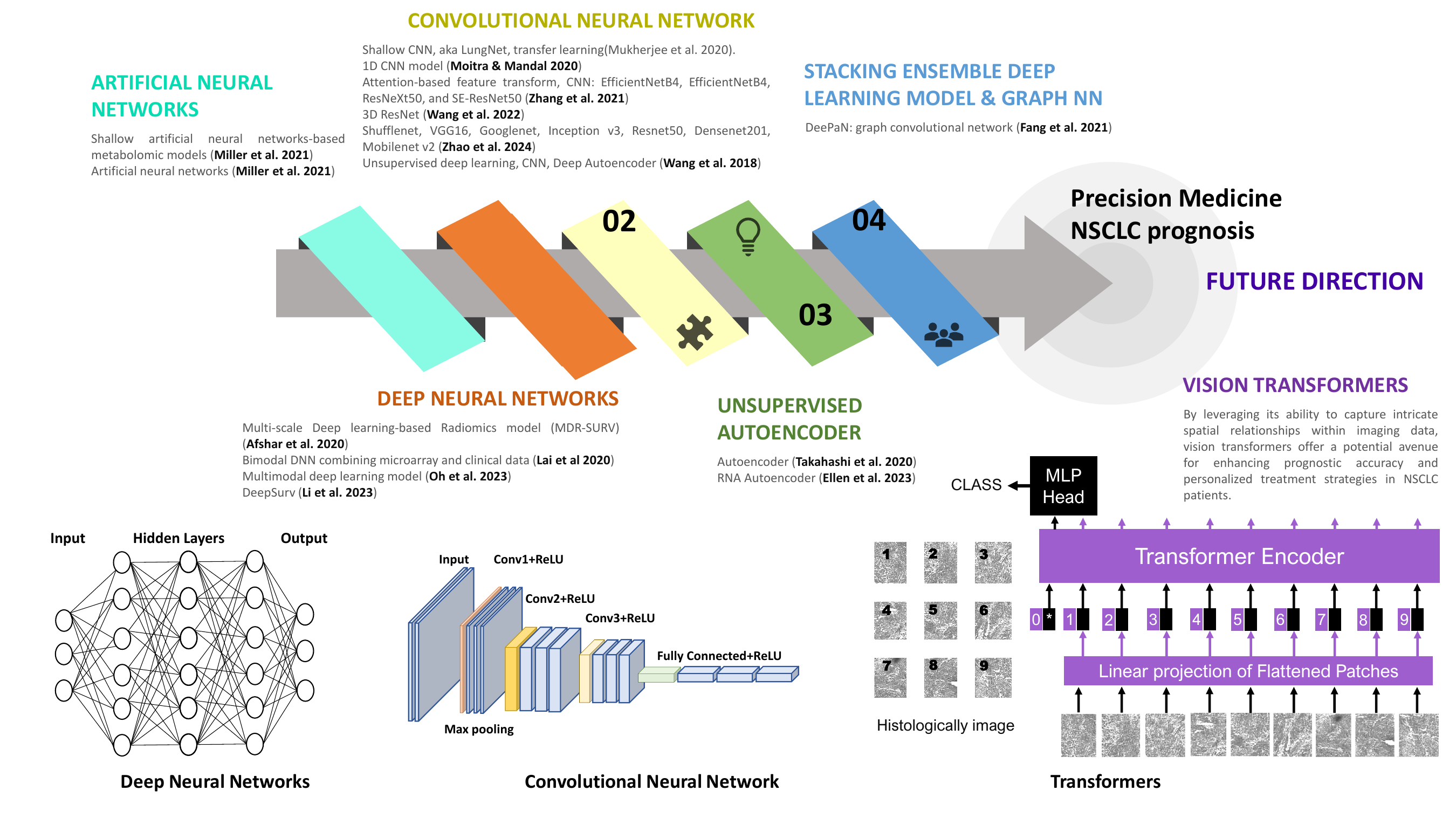}
\end{center}
   \caption{Deep learning has demonstrated notable applications in NSCLC prognosis, including the integration of genomic, multiomics, pathological, and imaging data for survival prediction. Future research in this domain aims to refine deep learning models, enhance interpretability, and incorporate real-time clinical data to further improve prognostic accuracy and guide personalized treatment strategies for NSCLC patients. The future direction of deep neural networks entails the development of more sophisticated models and the integration of attention mechanisms and transformers for improved models.}
\label{fig3}
\end{figure}

Deep neural networks revolutionized conventional machine learning and its effect led to improvements in NSCLC prognosis (see Figure \ref{fig3} and Table 1). A challenge of lung cancer OS analysis using CT images was addressed through utilizing unsupervised deep learning, residual convolutional autoencoder, to leverage unlabeled data for survival analysis, demonstrating superior performance compared to handcrafted features ~\cite{r400}. LungNet, a shallow convolutional neural network (CNN), served as a non-invasive predictor for NSCLC prognosis (and diagnosis), showcasing the potential of CNNs in interpreting CT images for lung cancer stratification and prognostication ~\cite{r401}. Lightweight models, i.e. 1D CNN model ~\cite{r402}, offer efficient NSCLC detection, consuming fewer resources, and time than traditional models, potentially serving as a decision support system for oncologists and radiologists. Deep neural network models were also developed for accurate OS prediction in NSCLC patients, combining gene expression and clinical data, suggesting its potential as a valuable tool for developing personalized therapies and advancing precision medicine in NSCLC ~\cite{r404}. 

  % ------------------------------------ Figure 3 --------------------
  % ------------------------------------ Figure 3 --------------------

% ================================  Table 1 ========================================
\begin{table*}[t]
\begin{center}
\caption{Summary of Deep learning and AI-based models for NSCLC prognosis} \vspace{-0.in}
\label{notations}
\includegraphics[width=0.87\linewidth]{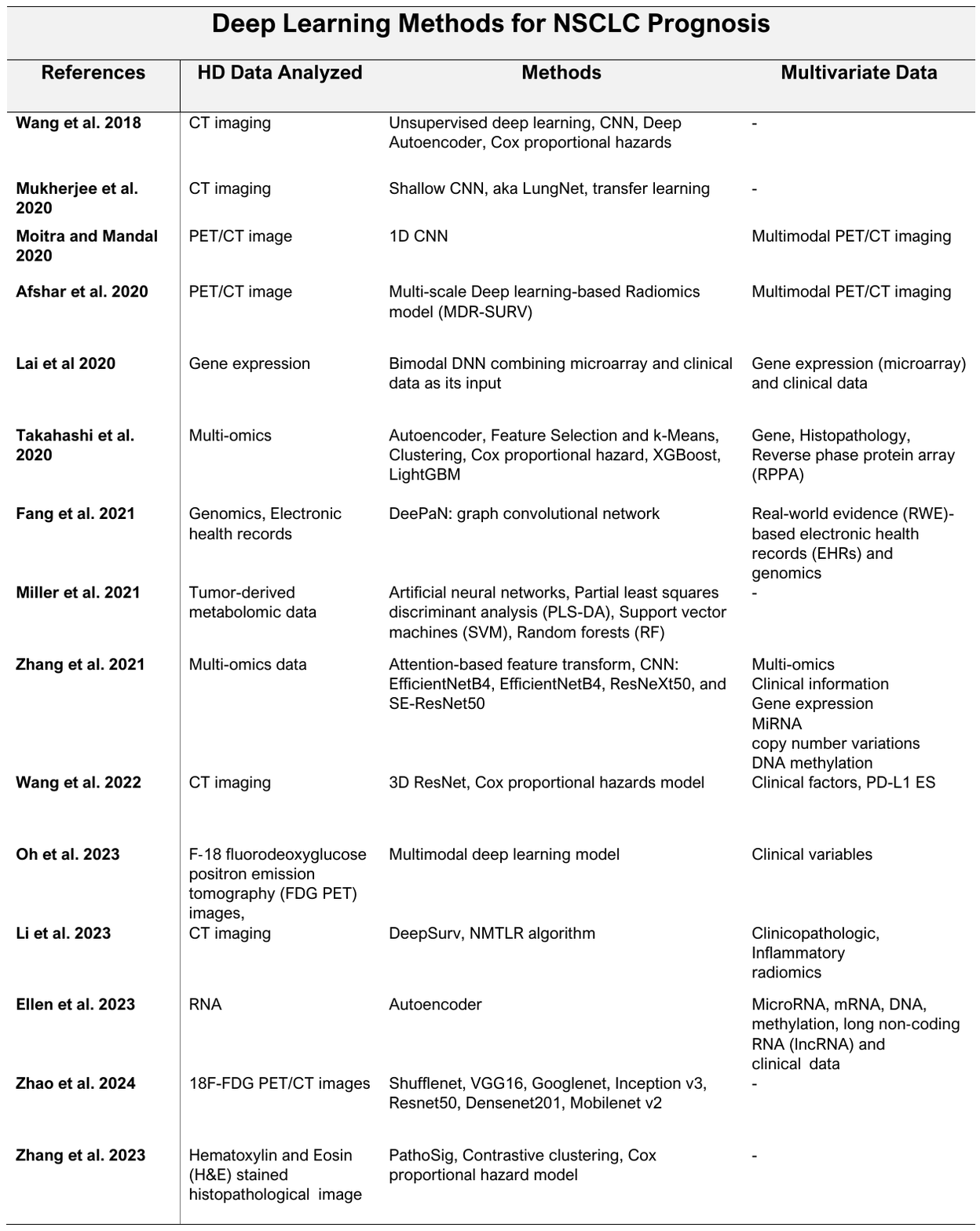} 
\end{center} \vspace{-0.in}
\label{table1}
\end{table*}
% ===================================================================================

A stacking ensemble deep learning model (1D-CNN) incorporating the LASSO technique for feature selection, demonstrated high performance metrics in RNA-seq analysis which led to the classification of five distinct cancer types ~\cite{r412}. Deep learning-based multi-model ensemble, employing a two-stage model with the DESeq technique for S-fold cross-validation, demonstrated remarkable accuracy which prevents overfitting and enhances predictive accuracy in RNA-seq analyses ~\cite{r411}. Multimodal integration strategies utilizing microRNA, mRNA, DNA methylation, long non-coding RNA (lncRNA), and clinical data showed enhanced survival prediction over single-modality models while employing denoise autoencoders for data compression and integration ~\cite{r420}. Utilization of k-fold cross-validation in deep neural network models employing Focal Loss may affect the model's reliability, while KL divergence served as the model features and efficacy of their approach in RNA-seq analysis ~\cite{r413}. Integrating clinical information and multi-omics data (gene expression, miRNA, copy number variations, and DNA methylation) for the deep neural network may result in the transformation of the given features which leads to improved prognostic prediction for NSCLC ~\cite{r408}. 

Insights into the intricate landscape of lung cancer subtypes, based on 450 K DNA methylation data, also suggested employing an unsupervised deep learning approach utilizing Variational Autoencoders (VAEs) to achieve high precision in classification ~\cite{r415}. Adjusting prediction accuracy using the CNN framework operating on protein-protein interaction (PPIs) networks and gene expression profiles suggested marginal improvements compared to conventional algorithms emphasizing the delicate balance between model complexity and performance gains ~\cite{r414}. Utilizing multi-omics analysis resulted in surviving groups (subtypes), independent of histopathological classification, and on reverse phase protein array (RPPA) data, which led to improved prognosis in lung cancer patients ~\cite{r405}. 

DeePaN, deep patient graph convolutional network, revealed distinct subgroups with significantly different median survival post-IO therapy, outperforming non-graph-based methods and offering insights into potential immune-oncological (IO) biomarkers, such as tumor mutation burden (TMB), and therapeutic considerations like KRAS mutations and blood monocyte count ~\cite{r406}. Shallow artificial neural networks-based metabolomic models also may improve the prediction of staging and chemotherapy response ~\cite{r407}. 

Integrating clinicopathological, inflammatory, and radiomics features through multidimensional deep neural network models, i.e. using DeepSurv and NMTLR algorithms ~\cite{r417} or ~\cite{r305,r306}, demonstrated enhanced predictive accuracy, aiding in personalized prognosis and guiding optimal treatment strategies, especially for patients identified in the high-risk group with significantly low PFS and OS. An extension of deep-learning analysis using Hematoxylin and Eosin-stained histopathological images resulted in identifying 50 histomorphological phenotype clusters (HPCs) as pathomic features to improve prognostic precision and therapeutic assessment in SCLC patients ~\cite{r419}.

The imaging-based applications of deep neural networks showed significant improvements in all aspects of lung cancer prognosis. These methods used Multi-scale Deep learning-based Radiomics 1model, MDR-SURV ~\cite{r403}, iterative sure independence screening (ISIS) scheme ~\cite{r409}, 3D ResNet on CT imaging and measurement of programmed death-ligand 1 (PD-L1) expression signature (ES) ~\cite{r410} or 3d ResNet on F-18 fluorodeoxyglucose PET (FDG PET) images with clinical data ~\cite{r416}, and an end-to-end deep learning model, Mobilenet v2, leveraging PET/CT images.

\section{Conclusions}

The synergistic integration of AI methodologies, spanning different aspects of prognostic analysis such as genomic, multi-omics, pathological, and imaging biomarkers, alongside sophisticated AI models, i.e., deep learning approaches, represents a notable advancement in the field of NSCLC prognosis. Through the intricate analysis and synthesis of multifaceted data, AI-driven models have revolutionized survival prediction, offering profound insights into the disease's complexity. This transformative paradigm not only enhances prognostic accuracy but also facilitates the tailoring of personalized treatment regimens, heralding a new era of precision medicine in NSCLC.
% \label{sec:1}
\section*{ACKNOWLEDGMENTS}
This work was supported, in part, by the University System of Maryland's William E. Kirwan Center for Academic Innovation grant 2022-2023 and TLTC T-02 (2024) funding initiative for incorporating artificial intelligence and machine learning. 

% See figure \ref{fig1} . . .

% \section*{Acknowledgments}
% This was was supported in part by......

%Bibliography
% \bibliographystyle{unsrt}  
% \bibliography{references} 

\begin{thebibliography}{}
   \bibitem{rr1}Govindan, R., Page, N., Morgensztern, D., Read, W., Tierney, R., Vlahiotis, A., Spitznagel, E. \& Piccirillo, J. Changing epidemiology of small-cell lung cancer in the United States over the last 30 years: analysis of the surveillance, epidemiologic, and end results database. {\em Journal Of Clinical Oncology}. \textbf{24}, 4539-4544 (2006)
\bibitem{rr2}Ferlay, J., Colombet, M., Soerjomataram, I., Parkin, D., Piñeros, M., Znaor, A. \& Bray, F. Cancer statistics for the year 2020: An overview. {\em International Journal Of Cancer}. \textbf{149}, 778-789 (2021)
\bibitem{rr3}Han, J., Jentzen, A. \& E, W. Solving high-dimensional partial differential equations using deep learning. {\em Proceedings Of The National Academy Of Sciences}. \textbf{115}, 8505-8510 (2018)
\bibitem{r1}Singhal, S., Miller, D., Ramalingam, S. \& Sun, S. Gene expression profiling of non-small cell lung cancer. {\em Lung Cancer}. \textbf{60}, 313-324 (2008)
\bibitem{r2}Heineman, D., Daniels, J. \& Schreurs, W. Clinical staging of NSCLC: current evidence and implications for adjuvant chemotherapy. {\em Therapeutic Advances In Medical Oncology}. \textbf{9}, 599-609 (2017)
\bibitem{r3}Petty, R., Nicolson, M., Kerr, K., Collie-Duguid, E. \& Murray, G. Gene expression profiling in non-small cell lung cancer: from molecular mechanisms to clinical application. {\em Clinical Cancer Research}. \textbf{10}, 3237-3248 (2004)
\bibitem{r4}Beer, D., Kardia, S., Huang, C., Giordano, T., Levin, A., Misek, D., Lin, L., Chen, G., Gharib, T., Thomas, D. \& Others Gene-expression profiles predict survival of patients with lung adenocarcinoma. {\em Nature Medicine}. \textbf{8}, 816-824 (2002)
\bibitem{r5}Kratz, J., Haro, G., Cook, N., He, J., Van Den Eeden, S., Woodard, G., Gubens, M., Jahan, T., Jones, K., Kim, I. \& Others Incorporation of a molecular prognostic classifier improves conventional non–small cell lung cancer staging. {\em Journal Of Thoracic Oncology}. \textbf{14}, 1223-1232 (2019)
\bibitem{r6}Välk, K., Vooder, T., Kolde, R., Reintam, M., Petzold, C., Vilo, J. \& Metspalu, A. Gene expression profiles of non-small cell lung cancer: survival prediction and new biomarkers. {\em Oncology}. \textbf{79}, 283-292 (2011)
\bibitem{r7}Bhattacharjee, A., Richards, W., Staunton, J., Li, C., Monti, S., Vasa, P., Ladd, C., Beheshti, J., Bueno, R., Gillette, M. \& Others Classification of human lung carcinomas by mRNA expression profiling reveals distinct adenocarcinoma subclasses. {\em Proceedings Of The National Academy Of Sciences}. \textbf{98}, 13790-13795 (2001)
\bibitem{r8}Guo, L., Ma, Y., Ward, R., Castranova, V., Shi, X. \& Qian, Y. Constructing molecular classifiers for the accurate prognosis of lung adenocarcinoma. {\em Clinical Cancer Research}. \textbf{12}, 3344-3354 (2006)
\bibitem{r9}Kikuchi, T., Daigo, Y., Katagiri, T., Tsunoda, T., Okada, K., Kakiuchi, S., Zembutsu, H., Furukawa, Y., Kawamura, M., Kobayashi, K. \& Others Expression profiles of non-small cell lung cancers on cDNA microarrays: identification of genes for prediction of lymph-node metastasis and sensitivity to anti-cancer drugs. {\em Oncogene}. \textbf{22}, 2192-2205 (2003)
\bibitem{r10}Petty, R., Kerr, K., Murray, G., Nicolson, M., Rooney, P., Bissett, D. \& Collie-Duguid, E. Tumor transcriptome reveals the predictive and prognostic impact of lysosomal protease inhibitors in non-small-cell lung cancer. {\em J Clin Oncol}. \textbf{24}, 1729-1744 (2006)
\bibitem{r11}Casagrande, G., Silva, M., Reis, R. \& Leal, L. Liquid biopsy for lung cancer: up-to-date and perspectives for screening programs. {\em International Journal Of Molecular Sciences}. \textbf{24}, 2505 (2023)
\bibitem{r12}Imamura, F., Uchida, J., Kukita, Y., Kumagai, T., Nishino, K., Inoue, T., Kimura, M., Oba, S. \& Kato, K. Monitoring of treatment responses and clonal evolution of tumor cells by circulating tumor DNA of heterogeneous mutant EGFR genes in lung cancer. {\em Lung Cancer}. \textbf{94} pp. 68-73 (2016)
\bibitem{r13}Pender, A., Garcia-Murillas, I., Rana, S., Cutts, R., Kelly, G., Fenwick, K., Kozarewa, I., Castro, D., Bhosle, J., O’Brien, M. \& Others Efficient genotyping of KRAS mutant non-small cell lung cancer using a multiplexed droplet digital PCR approach. {\em PloS One}. \textbf{10}, e0139074 (2015)
\bibitem{r14}Ye, Q., Hickey, J., Summers, K., Falatovich, B., Gencheva, M., Eubank, T., Ivanov, A. \& Guo, N. Multi-Omics Immune Interaction Networks in Lung Cancer Tumorigenesis, Proliferation, and Survival. {\em International Journal Of Molecular Sciences}. \textbf{23}, 14978 (2022)
\bibitem{r15}Aerts, H., Grossmann, P., Tan, Y., Oxnard, G., Rizvi, N., Schwartz, L. \& Zhao, B. Defining a radiomic response phenotype: a pilot study using targeted therapy in NSCLC. {\em Scientific Reports}. \textbf{6}, 33860 (2016)
\bibitem{r16}Chen, W., Qiao, X., Yin, S., Zhang, X. \& Xu, X. Integrating radiomics with genomics for non-small cell lung cancer survival analysis. {\em Journal Of Oncology}. \textbf{2022} (2022)
\bibitem{r300}Yu, K., Zhang, C., Berry, G., Altman, R., Ré, C., Rubin, D. \& Snyder, M. Predicting non-small cell lung cancer prognosis by fully automated microscopic pathology image features. {\em Nature Communications}. \textbf{7}, 12474 (2016)
\bibitem{r301}Coroller, T., Agrawal, V., Huynh, E., Narayan, V., Lee, S., Mak, R. \& Aerts, H. Radiomic-based pathological response prediction from primary tumors and lymph nodes in NSCLC. {\em Journal Of Thoracic Oncology}. \textbf{12}, 467-476 (2017)
\bibitem{r302}Luo, X., Yin, S., Yang, L., Fujimoto, J., Yang, Y., Moran, C., Kalhor, N., Weissferdt, A., Xie, Y., Gazdar, A. \& Others Development and Validation of a pathology Image Analysis-based predictive Model for Lung Adenocarcinoma prognosis-A Multi-cohort study. {\em Scientific Reports}. \textbf{9}, 6886 (2019)
\bibitem{r303}Widmaier, M., Wiestler, T., Walker, J., Barker, C., Scott, M., Sekhavati, F., Budco, A., Schneider, K., Segerer, F., Steele, K. \& Others Comparison of continuous measures across diagnostic PD-L1 assays in non-small cell lung cancer using automated image analysis. {\em Modern Pathology}. \textbf{33}, 380-390 (2020)
\bibitem{r304}Zhang, S., Fan, Y., Zhong, T. \& Ma, S. Histopathological imaging features-versus molecular measurements-based cancer prognosis modeling. {\em Scientific Reports}. \textbf{10}, 15030 (2020)
\bibitem{r305}Kanavati, F., Toyokawa, G., Momosaki, S., Takeoka, H., Okamoto, M., Yamazaki, K., Takeo, S., Iizuka, O. \& Tsuneki, M. A deep learning model for the classification of indeterminate lung carcinoma in biopsy whole slide images. {\em Scientific Reports}. \textbf{11}, 8110 (2021)
\bibitem{r306}Diao, J., Wang, J., Chui, W., Mountain, V., Gullapally, S., Srinivasan, R., Mitchell, R., Glass, B., Hoffman, S., Rao, S. \& Others Human-interpretable image features derived from densely mapped cancer pathology slides predict diverse molecular phenotypes. {\em Nature Communications}. \textbf{12}, 1613 (2021)
\bibitem{r307}Khodabakhshi, Z., Mostafaei, S., Arabi, H., Oveisi, M., Shiri, I. \& Zaidi, H. Non-small cell lung carcinoma histopathological subtype phenotyping using high-dimensional multinomial multiclass CT radiomics signature. {\em Computers In Biology And Medicine}. \textbf{136} pp. 104752 (2021)
\bibitem{r98}Limkin, E., Sun, R., Dercle, L., Zacharaki, E., Robert, C., Reuzé, S., Schernberg, A., Paragios, N., Deutsch, E. \& Ferté, C. Promises and challenges for the implementation of computational medical imaging (radiomics) in oncology. {\em Annals Of Oncology}. \textbf{28}, 1191-1206 (2017)
\bibitem{r99}Thawani, R., McLane, M., Beig, N., Ghose, S., Prasanna, P., Velcheti, V. \& Madabhushi, A. Radiomics and radiogenomics in lung cancer: a review for the clinician. {\em Lung Cancer}. \textbf{115} pp. 34-41 (2018)
\bibitem{r99+1}Jayasurya, K., Fung, G., Yu, S., Dehing-Oberije, C., De Ruysscher, D., Hope, A., De Neve, W., Lievens, Y., Lambin, P. \& Dekker, A. Comparison of Bayesian network and support vector machine models for two-year survival prediction in lung cancer patients treated with radiotherapy. {\em Medical Physics}. \textbf{37}, 1401-1407 (2010)
\bibitem{r99+2}Sun, T., Wang, J., Li, X., Lv, P., Liu, F., Luo, Y., Gao, Q., Zhu, H. \& Guo, X. Comparative evaluation of support vector machines for computer aided diagnosis of lung cancer in CT based on a multi-dimensional data set. {\em Computer Methods And Programs In Biomedicine}. \textbf{111}, 519-524 (2013)
\bibitem{r99+3}Hyun, S., Ahn, M., Koh, Y. \& Lee, S. A machine-learning approach using PET-based radiomics to predict the histological subtypes of lung cancer. {\em Clinical Nuclear Medicine}. \textbf{44}, 956-960 (2019)
\bibitem{r100}Wang, H., Xing, F., Su, H., Stromberg, A. \& Yang, L. Novel image markers for non-small cell lung cancer classification and survival prediction. {\em BMC Bioinformatics}. \textbf{15} pp. 1-12 (2014)
\bibitem{r101}Hawkins, S., Wang, H., Liu, Y., Garcia, A., Stringfield, O., Krewer, H., Li, Q., Cherezov, D., Gatenby, R., Balagurunathan, Y. \& Others Predicting malignant nodules from screening CT scans. {\em Journal Of Thoracic Oncology}. \textbf{11}, 2120-2128 (2016)
\bibitem{r102}Liu, Y., Kim, J., Balagurunathan, Y., Li, Q., Garcia, A., Stringfield, O., Ye, Z. \& Gillies, R. Radiomic features are associated with EGFR mutation status in lung adenocarcinomas. {\em Clinical Lung Cancer}. \textbf{17}, 441-448 (2016)
\bibitem{r103}Shen, C., Liu, Z., Guan, M., Song, J., Lian, Y., Wang, S., Tang, Z., Dong, D., Kong, L., Wang, M. \& Others 2D and 3D CT radiomics features prognostic performance comparison in non-small cell lung cancer. {\em Translational Oncology}. \textbf{10}, 886-894 (2017)
\bibitem{r104}Fave, X., Zhang, L., Yang, J., Mackin, D., Balter, P., Gomez, D., Followill, D., Jones, A., Stingo, F., Liao, Z. \& Mohan, R. Delta-radiomics features for the prediction of patient outcomes in non–small cell lung cancer. {\em Scientific Reports}. \textbf{7}, 588 (2017)
\bibitem{r105}Song, S., Park, H., Lee, G., Lee, H., Sohn, I., Kim, H., Lee, S., Jeong, J., Kim, J., Lee, K. \& Others Imaging phenotyping using radiomics to predict micropapillary pattern within lung adenocarcinoma. {\em Journal Of Thoracic Oncology}. \textbf{12}, 624-632 (2017)
\bibitem{r106}Ganeshan, B., Goh, V., Mandeville, H., Ng, Q., Hoskin, P. \& Miles, K. Non–small cell lung cancer: histopathologic correlates for texture parameters at CT. {\em Radiology}. \textbf{266}, 326-336 (2013)
\bibitem{r107}Zhang, Y., Oikonomou, A., Wong, A., Haider, M. \& Khalvati, F. Radiomics-based prognosis analysis for non-small cell lung cancer. {\em Scientific Reports}. \textbf{7}, 46349 (2017)
\bibitem{r108}Saad, M. \& Choi, T. Computer-assisted subtyping and prognosis for non-small cell lung cancer patients with unresectable tumor. {\em Computerized Medical Imaging And Graphics}. \textbf{67} pp. 1-8 (2018)
\bibitem{r109}Zhou, H., Dong, D., Chen, B., Fang, M., Cheng, Y., Gan, Y., Zhang, R., Zhang, L., Zang, Y., Liu, Z. \& Others Diagnosis of distant metastasis of lung cancer: based on clinical and radiomic features. {\em Translational Oncology}. \textbf{11}, 31-36 (2018)
\bibitem{r110}Sun, W., Jiang, M., Dang, J., Chang, P. \& Yin, F. Effect of machine learning methods on predicting NSCLC overall survival time based on Radiomics analysis. {\em Radiation Oncology}. \textbf{13}, 1-8 (2018)
\bibitem{r111}Bianconi, F., Fravolini, M., Bello-Cerezo, R., Minestrini, M., Scialpi, M. \& Palumbo, B. Evaluation of shape and textural features from CT as prognostic biomarkers in non-small cell lung cancer. {\em Anticancer Research}. \textbf{38}, 2155-2160 (2018)
\bibitem{r112}He, B., Zhao, W., Pi, J., Han, D., Jiang, Y., Zhang, Z. \& Zhao, W. A biomarker basing on radiomics for the prediction of overall survival in non–small cell lung cancer patients. {\em Respiratory Research}. \textbf{19} pp. 1-8 (2018)
\bibitem{r113}Zhang, L., Chen, B., Liu, X., Song, J., Fang, M., Hu, C., Dong, D., Li, W. \& Tian, J. Quantitative biomarkers for prediction of epidermal growth factor receptor mutation in non-small cell lung cancer. {\em Translational Oncology}. \textbf{11}, 94-101 (2018)
\bibitem{r114}Junior, J., Koenigkam-Santos, M., Cipriano, F., Fabro, A. \& Azevedo-Marques, P. Radiomics-based features for pattern recognition of lung cancer histopathology and metastases. {\em Computer Methods And Programs In Biomedicine}. \textbf{159} pp. 23-30 (2018)
\bibitem{r115}Yang, L., Yang, J., Zhou, X., Huang, L., Zhao, W., Wang, T., Zhuang, J. \& Tian, J. Development of a radiomics nomogram based on the 2D and 3D CT features to predict the survival of non-small cell lung cancer patients. {\em European Radiology}. \textbf{29} pp. 2196-2206 (2019)
\bibitem{r116}Farchione, A., Larici, A., Masciocchi, C., Cicchetti, G., Congedo, M., Franchi, P., Gatta, R., Lo Cicero, S., Valentini, V., Bonomo, L. \& Others Exploring technical issues in personalized medicine: NSCLC survival prediction by quantitative image analysis—usefulness of density correction of volumetric CT data. {\em La Radiologia Medica}. \textbf{125} pp. 625-635 (2020)
\bibitem{r117}Pinheiro, G., Pereira, T., Dias, C., Freitas, C., Hespanhol, V., Costa, J., Cunha, A. \& Oliveira, H. Identifying relationships between imaging phenotypes and lung cancer-related mutation status: EGFR and KRAS. {\em Scientific Reports}. \textbf{10}, 3625 (2020)
\bibitem{r118}Yousefi, B., Jahani, N., LaRiviere, M., Cohen, E., Hsieh, M., Luna, J., Chitalia, R., Thompson, J., Carpenter, E., Katz, S. \& Others Correlative hierarchical clustering-based low-rank dimensionality reduction of radiomics-driven phenotype in non-small cell lung cancer. {\em Medical Imaging 2019: Imaging Informatics For Healthcare, Research, And Applications}. \textbf{10954} pp. 278-285 (2019)
\bibitem{r119}Yousefi, B., LaRiviere, M., Cohen, E., Buckingham, T., Yee, S., Black, T., Chien, A., Noël, P., Hwang, W., Katz, S. \& Others Combining radiomic phenotypes of non-small cell lung cancer with liquid biopsy data may improve prediction of response to EGFR inhibitors. {\em Scientific Reports}. \textbf{11}, 9984 (2021)
\bibitem{r120}Hershman, M., Yousefi, B., Serletti, L., Galperin-Aizenberg, M., Roshkovan, L., Luna, J., Thompson, J., Aggarwal, C., Carpenter, E., Kontos, D. \& Others Impact of interobserver variability in manual segmentation of non-small cell lung cancer (NSCLC) applying low-rank radiomic representation on computed tomography. {\em Cancers}. \textbf{13}, 5985 (2021)
\bibitem{r121}Horng, H., Singh, A., Yousefi, B., Cohen, E., Haghighi, B., Katz, S., Noël, P., Shinohara, R. \& Kontos, D. Generalized ComBat harmonization methods for radiomic features with multi-modal distributions and multiple batch effects. {\em Scientific Reports}. \textbf{12}, 4493 (2022)
\bibitem{r122}Horng, H., Singh, A., Yousefi, B., Cohen, E., Haghighi, B., Katz, S., Noël, P., Kontos, D. \& Shinohara, R. Improved generalized ComBat methods for harmonization of radiomic features. {\em Scientific Reports}. \textbf{12}, 19009 (2022)
\bibitem{r123}Lockhart, J., Ackerman, H., Lee, K., Abdalah, M., Davis, A., Hackel, N., Boyle, T., Saller, J., Keske, A., Hänggi, K. \& Others Grading of lung adenocarcinomas with simultaneous segmentation by artificial intelligence (GLASS-AI). {\em NPJ Precision Oncology}. \textbf{7}, 68 (2023)
\bibitem{r199}Cook, G., Azad, G., Owczarczyk, K., Siddique, M. \& Goh, V. Challenges and promises of PET radiomics. {\em International Journal Of Radiation Oncology* Biology* Physics}. \textbf{102}, 1083-1089 (2018)
\bibitem{r200}Oliver, J., Budzevich, M., Zhang, G., Dilling, T., Latifi, K. \& Moros, E. Variability of image features computed from conventional and respiratory-gated PET/CT images of lung cancer. {\em Translational Oncology}. \textbf{8}, 524-534 (2015)
\bibitem{r201}Gevaert, O., Xu, J., Hoang, C., Leung, A., Xu, Y., Quon, A., Rubin, D., Napel, S. \& Plevritis, S. Non–small cell lung cancer: identifying prognostic imaging biomarkers by leveraging public gene expression microarray data—methods and preliminary results. {\em Radiology}. \textbf{264}, 387-396 (2012)
\bibitem{r400}Wang, S., Liu, Z., Chen, X., Zhu, Y., Zhou, H., Tang, Z., Wei, W., Dong, D., Wang, M. \& Tian, J. Unsupervised deep learning features for lung cancer overall survival analysis. {\em 2018 40th Annual International Conference Of The IEEE Engineering In Medicine And Biology Society (EMBC)}. pp. 2583-2586 (2018)
\bibitem{r401}Mukherjee, P., Zhou, M., Lee, E., Schicht, A., Balagurunathan, Y., Napel, S., Gillies, R., Wong, S., Thieme, A., Leung, A. \& Others A shallow convolutional neural network predicts prognosis of lung cancer patients in multi-institutional computed tomography image datasets. {\em Nature Machine Intelligence}. \textbf{2}, 274-282 (2020)
\bibitem{r402}Moitra, D. \& Mandal, R. Classification of non-small cell lung cancer using one-dimensional convolutional neural network. {\em Expert Systems With Applications}. \textbf{159} pp. 113564 (2020)
\bibitem{r403}Afshar, P., Oikonomou, A., Plataniotis, K. \& Mohammadi, A. MDR-SURV: a multi-scale deep learning-based radiomics for survival prediction in pulmonary malignancies. {\em ICASSP 2020-2020 IEEE International Conference On Acoustics, Speech And Signal Processing (ICASSP)}. pp. 2013-2017 (2020)
\bibitem{r404}Lai, Y., Chen, W., Hsu, T., Lin, C., Tsao, Y. \& Wu, S. Overall survival prediction of non-small cell lung cancer by integrating microarray and clinical data with deep learning. {\em Scientific Reports}. \textbf{10}, 4679 (2020)
\bibitem{r405}Takahashi, S., Asada, K., Takasawa, K., Shimoyama, R., Sakai, A., Bolatkan, A., Shinkai, N., Kobayashi, K., Komatsu, M., Kaneko, S. \& Others Predicting deep learning based multi-omics parallel integration survival subtypes in lung cancer using reverse phase protein array data. {\em Biomolecules}. \textbf{10}, 1460 (2020)
\bibitem{r406}Fang, C., Xu, D., Su, J., Dry, J. \& Linghu, B. DeePaN: dee p pa tient graph convolutional n etwork integrating clinico-genomic evidence to stratify lung cancers for immunotherapy. {\em NPJ Digital Medicine}. \textbf{4}, 14 (2021)
\bibitem{r407}Miller, H., Yin, X., Smith, S., Hu, X., Zhang, X., Yan, J., Miller, D., Berkel, V. \& Frieboes, H. Evaluation of disease staging and chemotherapeutic response in non-small cell lung cancer from patient tumor-derived metabolomic data. {\em Lung Cancer}. \textbf{156} pp. 20-30 (2021)
\bibitem{r408}Zhang, Z., Xu, F., Jiang, H. \& Chen, Z. Prognostic Prediction for Non-small-Cell Lung Cancer Based on Deep Neural Network and Multimodal Data. {\em Intelligent Computing Theories And Application: 17th International Conference, ICIC 2021, Shenzhen, China, August 12–15, 2021, Proceedings, Part III 17}. pp. 549-560 (2021)
\bibitem{r409}Bhattacharjee, A., Dey, J. \& Kumari, P. A combined iterative sure independence screening and Cox proportional hazard model for extracting and analyzing prognostic biomarkers of adenocarcinoma lung cancer. {\em Healthcare Analytics}. \textbf{2} pp. 100108 (2022)
\bibitem{r410}Wang, C., Ma, J., Shao, J., Zhang, S., Li, J., Yan, J., Zhao, Z., Bai, C., Yu, Y. \& Li, W. Non-invasive measurement using deep learning algorithm based on multi-source features fusion to predict PD-L1 expression and survival in NSCLC. {\em Frontiers In Immunology}. \textbf{13} pp. 828560 (2022)
\bibitem{r411}Xiao, Y., Wu, J., Lin, Z. \& Zhao, X. A deep learning-based multi-model ensemble method for cancer prediction. {\em Computer Methods And Programs In Biomedicine}. \textbf{153} pp. 1-9 (2018)
\bibitem{r412}Mohammed, M., Mwambi, H., Mboya, I., Elbashir, M. \& Omolo, B. A stacking ensemble deep learning approach to cancer type classification based on TCGA data. {\em Scientific Reports}. \textbf{11}, 15626 (2021)
\bibitem{r413}Liu, S. \& Yao, W. Prediction of lung cancer using gene expression and deep learning with KL divergence gene selection. {\em BMC Bioinformatics}. \textbf{23}, 175 (2022)
\bibitem{r414}Matsubara, T., Ochiai, T., Hayashida, M., Akutsu, T. \& Nacher, J. Convolutional neural network approach to lung cancer classification integrating protein interaction network and gene expression profiles. {\em Journal Of Bioinformatics And Computational Biology}. \textbf{17}, 1940007 (2019)
\bibitem{r415}Wang, Z. \& Wang, Y. Extracting a biologically latent space of lung cancer epigenetics with variational autoencoders. {\em BMC Bioinformatics}. \textbf{20}, 1-7 (2019)
\bibitem{r416}Oh, S., Kang, S., Oh, I. \& Kim, M. Deep learning model integrating positron emission tomography and clinical data for prognosis prediction in non-small cell lung cancer patients. {\em BMC Bioinformatics}. \textbf{24}, 1-13 (2023)
\bibitem{r417}Li, B., Yang, L., Jiang, C., Yao, Y., Li, H., Cheng, S., Zou, B., Fan, B. \& Wang, L. Integrated multi-dimensional deep neural network model improves prognosis prediction of advanced NSCLC patients receiving bevacizumab. {\em Frontiers In Oncology}. \textbf{13} pp. 1052147 (2023)
\bibitem{r418}Zhao, H., Su, Y., Lyu, Z., Tian, L., Xu, P., Lin, L., Han, W. \& Fu, P. Non-invasively Discriminating the Pathological Subtypes of Non-small Cell Lung Cancer with Pretreatment 18F-FDG PET/CT Using Deep Learning. {\em Academic Radiology}. \textbf{31}, 35-45 (2024)
\bibitem{r419}Zhang, Y., Yang, Z., Chen, R., Zhu, Y., Liu, L., Dong, J., Zhang, Z., Sun, X., Ying, J., Lin, D. \& Others Histopathology images-based deep learning prediction of prognosis and therapeutic response in small cell lung cancer. {\em NPJ Digital Medicine}. \textbf{7}, 15 (2024)
\bibitem{r420}Ellen, J., Jacob, E., Nikolaou, N. \& Markuzon, N. Autoencoder-based multimodal prediction of non-small cell lung cancer survival. {\em Scientific Reports}. \textbf{13}, 15761 (2023)
\bibitem{r51}Gevaert, O., Xu, J., Hoang, C., Leung, A., Xu, Y., Quon, A., Rubin, D., Napel, S. \& Plevritis, S. Non–small cell lung cancer: identifying prognostic imaging biomarkers by leveraging public gene expression microarray data—methods and preliminary results. {\em Radiology}. \textbf{264}, 387-396 (2012)
 %
\end{thebibliography}

\end{document}